\documentclass[%
 amsmath,amssymb,
 aps,
 prx
]{revtex4-2}

\usepackage{graphicx}
\usepackage{dcolumn}
\usepackage{bm}
\usepackage{hyperref}

\usepackage{CJKutf8}
\newcommand{\japanese}[1]{\begin{CJK*}{UTF8}{min}#1\end{CJK*}}
\usepackage{siunitx}

\begin{document}
\title{Geospatial analysis of toponyms in geotagged social media posts}

\author{Takayuki Hiraoka}
\email{takayuki.hiraoka@aalto.fi}
\affiliation{Department of Computer Science, Aalto University, Espoo, Finland}
\author{Takashi Kirimura}
\affiliation{Department of Kyoto Studies, Kyoto Sangyo Univerisity, Kyoto, Japan}
\author{Naoya Fujiwara}
\affiliation{Graduate School of Information Sciences, Tohoku University, Sendai, Japan}
\affiliation{PRESTO, Japan Science and Technology Agency, Kawaguchi, Japan}
\affiliation{Institute of Industrial Science, The University of Tokyo, Tokyo, Japan}
\affiliation{Center for Spatial Information Science, The University of Tokyo, Kashiwa, Japan}

\begin{abstract}
Place names, or \emph{toponyms}, play an integral role in human representation and communication of geographic space. In particular, how people relate each toponym with particular locations in geographic space should be indicative of their spatial perception. Here, we make use of an extensive dataset of georeferenced social media posts, retrieved from Twitter, to perform a statistical analysis of the geographic distribution of toponyms and uncover the relationship between toponyms and geographic space. We show that the occurrence of toponyms is characterized by spatial inhomogeneity, giving rise to patterns that are distinct from the distribution of common nouns. Using simple models, we quantify the spatial specificity of toponym distributions and identify their core-periphery structures. In particular, we find that toponyms are used with a probability that decays as a power law with distance from the geographic center of their occurrence. Our findings highlight the potential of social media data to explore linguistic patterns in geographic space, paving the way for comprehensive analyses of human spatial representations. 
\end{abstract}
\maketitle

\section{Introduction}
When we speak or write about geographic spaces, we generally communicate them using the names of places, or toponyms. Although any point on Earth can be specified by geographic coordinates, one would rarely refer to a place in day-to-day conversation as, for example, ``\ang{35.7}N, \ang{139.7}E''; instead, we would usually represent it by a toponym, such as \emph{Tokyo}. The use of toponyms reflects the way we perceive and mentally structure geographic space. Unlike geographic coordinates, which are objective and unambiguous, the area a toponym refers to is often vague and difficult to define; even for the names of administratively defined areas (such as municipalities), how people use them in colloquial settings may have only a loose correspondence with the officially demarcated boundaries~\cite{Montello2003, Jones2008}. This however does not mean that the extent that each toponym denotes can be arbitrarily defined by individuals; since the main function of toponyms is to effectively communicate geographic information, there must be a shared consensus within the population that determines the areas they represent. 

The objective of this study is to understand such a spontaneous and collective relationship between geographic space and the use of toponyms. In order to quantitatively examine this relationship, one needs to collect large-scale data on the locations people associate with each toponym. The abundance of user-generated content online, especially on social media, offers a promising opportunity for this purpose. In particular, Twitter (rebranded to X in 2023), one of the major social media platforms, had provided free access to posts on the platform through APIs until 2023, allowing researchers to perform statistical analysis and find patterns and trends in the data. On Twitter, users could opt to attach to each of their post geolocation metadata (a \emph{geotag}) that represent the geographic coordinates of the GPS location of their device. These user-annotated geotags as well as the content of the posts allow us to explore the spatial dimension of user behavior and language use. For instance, by leveraging the fact that the set of vocabulary that appears in the text of posts varies according to the whereabouts of users, studies have found that toponyms in the text can be disambiguated~\cite{DeLozier2015} and the location of individual users can be identified~\cite{Cheng2010, Li2011}, although there are limitations to this approach at high spatial resolution~\cite{Hahmann2014}. When aggregated at the population level, the data can be used to study dialectal variation and language evolution, making it of interest to sociolinguistics and linguistic geography~\cite{Eisenstein2010, Huang2016, Abitbol2018, Louf2023American, Louf2023When, Morin2024}.

We note that the geotag attached to a post does not necessarily correspond to the place referred to in the text~\cite{Pavalanathan2015, Johnson2016}. This discrepancy can arise from inaccurate tagging~\cite{Oliveira2016}, but it can also be a manifestation of the geographic awareness and identity of the user, i.e., how users perceive and choose to represent their location or the place under discussion~\cite{Xu2013, Han2015, Arthur2019Human}. Users may (mis)represent their location for various reasons, such as privacy concerns, a desire to be associated with a particular place, or simply because they are referring to a location that is not their own.

In this study, we focus on understanding the geographic distribution of geotagged posts and the toponyms they contain. Our goal is to understand the collective, population-level knowledge about the relationship between toponyms and geographic locations rather than to identify patterns of toponym usage at the individual user level. Similar questions have been addressed in the literature: Hollenstein and Purves~\cite{Hollenstein2010} used user-annotated geotag data from the image hosting service Flickr to delineate city centers and neighborhoods; Hu and Janowicz~\cite{Hu2018Empirical} studied the names of points of interest (such as restaurants and shops) registered on Yelp, a user-generated local business review platform, in metropolitan areas in the United States. These studies were primarily focused on urban areas, and therefore the analysis was limited to a relatively small geographic scale. In addition, the results were rather descriptive and were not aimed at deriving general laws of toponymic occurrence. In contrast, we use a data-driven modeling approach to uncover the underlying principles governing the occurrence of toponyms of different granularities on a larger geographic scale, aiming to understand how these distributions reflect collective spatial cognition.

The rest of the paper is structured as follows. We start by introducing the geotagged Twitter dataset and how we collect, preprocess, and subsample it in the \emph{Data} section. In the \emph{Results} section, we first present the inhomogeneity of the geographic distribution of geotagged posts, and compare it to the population distributions. We then focus on the geographic distribution of individual toponyms. In the second subsection, we observe that the occurrence of each domestic toponym follows a characteristic pattern, hinting at its spatial specificity. To formalize this observation, we introduce a class of models called binomial models. In the third subsection, we use the simplest instance of this model to quantify the spatial specificity of each toponym. In the last subsection, we show that another variant of the binomial model, which we call the core-periphery model, reproduces the essential elements of toponym occurrence patterns despite its simplicity. In the \emph{Discussion and Conclusions} section, we discuss the implications of our work.

\section{Data}
We collected \num{395268777} geotagged Twitter posts from the Twitter API. These posts are annotated with coordinates within the bounding box of Japan (latitude between \ang{20.43}N and \ang{45.56}N and longitude between \ang{122.93}E and \ang{153.99}E) during the period from 1 February 2012 to 30 September 2018. Note that this bounding box also includes South and North Korea as well as parts of China and Russia. The data collection was conducted in accordance with Twitter's Terms of Service, which allow researchers to analyze and publish findings based on Twitter data, but prohibit the redistribution of raw data, such as the text or geotags of individual posts.

Data collection was followed by several preprocessing stages. First, we excluded posts made via location check-in services (e.g., Foursquare) as well as those generated by automated bots or manipulative users. Posts made through Foursquare are in specific text formats, such as ``I'm at [the name of a place/point of interest]'' or ``[post text] (@ [the name of a place/point of interest])''. When a user checks in to a place, they select the name of the place/point of interest, which may include toponyms, from a list of nearby places provided by Foursquare. However, users do not have independent control over the geographic coordinates tagged to the post; these coordinates are automatically determined by Foursquare's location database based on the name of the place chosen. Although these posts are created by personal users, they do not represent an organic association between toponyms and geographic coordinates. Including them in the analysis would bias the findings of this study.

In addition, the geotags attached to bot-generated posts may not reflect the geospatial behavior of individual users. Many non-personal accounts generate geotagged posts for various purposes, not necessarily with commercial or malicious intent;  for example, they may be bots that send out weather or traffic alerts~\cite{Guo2014}. Most of them are identifiable by the `source' metadata, which indicates the application or device used to make the post. Some accounts engage in more active geotag manipulation. Zhao and Sui~\cite{Zhao2017} developed a technique to detect manipulated geotags and concluded that such manipulations accounted for \qty{0.22}{\percent} of their sample, with even lower percentages among posts from official Twitter clients for iPhone and Android. 

Based on these considerations, we restricted our dataset to posts originating from official or general-use third-party mobile applications using the source metadata. This filtering process excluded \qty{29.87}{\percent} of the collected posts---\qty{20.06}{\percent} from location check-in services and \qty{9.81}{\percent} from other sources. A full list of the applications considered for inclusion in the dataset is available in Supporting Information. We further excluded posts tagged outside the geographic range of our study but accidentally included in the collected data; this reduces the sample size by an additional \qty{0.46}{\percent}.

Additionally, we found evidence that some users likely manipulated the geotags of their posts and assigned random geographic coordinates in unnatural rectangular bounding boxes that partly extend over sea areas (see Supporting Information). These posts can be characterized by containing a large number of mentions to other accounts, typically seven or more. Although not all posts with many mentions are necessarily manipulated, we conservatively excluded all posts with seven or more mentions. We confirmed that removing these posts, which account for \qty{0.06}{\percent} of the sample, did not significantly alter any of our findings.

After these filtering steps, the dataset consists of \num{275750003} posts. The geolocation metadata of each post is aggregated into grid cells based on the standard grid square system used in Japan's official spatial statistics. Each grid cell spans \ang{;;30} of latitude and \ang{;;45} of longitude, which is approximately a square with a side length of \SI{1}{km}, although the east-west width of a grid cell varies slightly with latitude. In total, \num{293444} grid cells contain a nonzero number of posts in the dataset.

As of 2019, Japan is divided into 47 prefectures as the first level of administrative division, and \num{1741} municipalities (cities, towns, villages, and twenty-three special wards of Tokyo) as the second level of division. In addition, some large cities, referred to as ``designated cities'', have administrative, non-autonomous subdivisions known as wards. Prefectures are sometimes grouped into seven to thirteen regions, although regions are not official administrative units, and the name and extent of each region can be ambiguous. 

\begin{figure}[tb]
    \centering
    \includegraphics[width=0.83\textwidth]{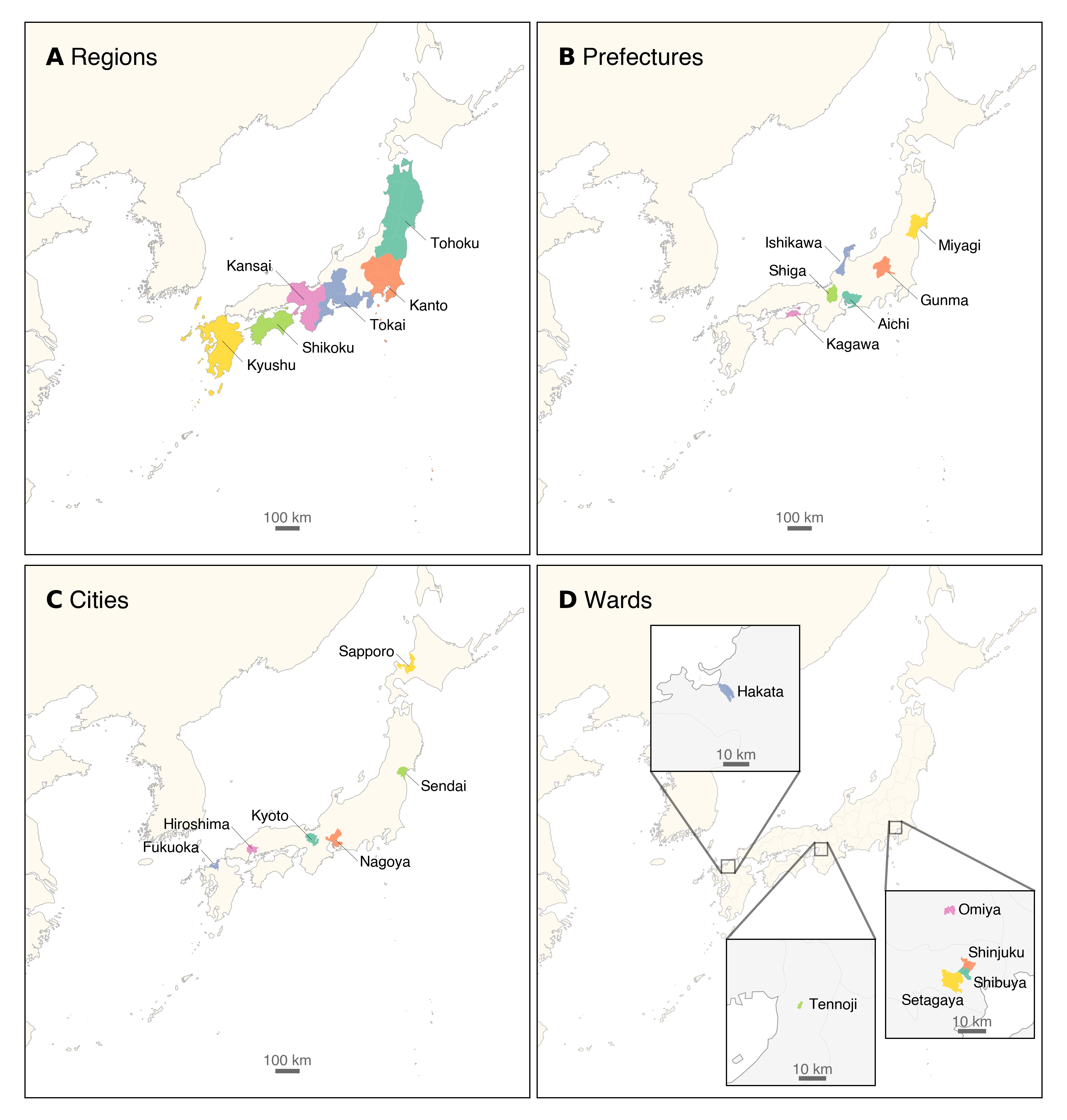}
    \caption{\textbf{Places in Japan denoted by the toponyms studied in this paper.} In this study, we sample 24 toponyms: the names of \textbf{(A)} six regions, \textbf{(B)} six prefectures, \textbf{(C)} six major cities, and \textbf{(D)} six wards (submetropolitan/submunicipal districts). The colored areas in each panel show the administratively defined geographic area (except for cities) denoted by each toponym. \textbf{(A)} The extent of each region is not uniquely defined. Here we show one of the commonly used classifications of regions. \textbf{(C)} The colored area shows the metropolitan employment area~\cite{Kanemoto2002, UEA}, which is considered to be more representative of urban activity than administratively defined city areas. Note that Kyoto, Hiroshima, and Fukuoka are used both as the names of the cities and as the names of the prefectures of which the cities are the capitals. \textbf{(D)} Prefectural boundaries are shown for visual guidance. Maps made with Natural Earth (\url{https://www.naturalearthdata.com/}).}
    \label{fig:toponym_map}
\end{figure}

From the full sample of the \num{276} million geotagged posts, we extract the subset of posts that contains each of 24 Japanese toponyms that refer to regions, prefectures, cities, and wards (special wards of Tokyo Metropolis and wards of designated cities) in Japan. The list of toponyms sampled in this work and the administrative areas they refer to can be found in Fig~\ref{fig:toponym_map}. 

Even if the text of a post contains a string that matches a toponym, it does not necessarily mean that the user is referring to the place it denotes. For instance, the name of the city Hiroshima is a substring of the name of another city Kitahiroshima, which is located over \SI{1200}{km} away. As a result, posts containing \emph{Hiroshima} include not only posts that refer to Hiroshima but also posts that mention Kitahiroshima. To separate the references to Hiroshima from the references to Kitahiroshima, we need to exclude the posts containing \emph{Kitahiroshima}. For each of the 24 toponyms we study in this paper, we discounted the posts that include the names of other regions, prefectures, cities with a population larger than \num{50000}, or wards that contain the toponym as a substring. We provide further details in Supporting Information.

In addition to these domestic toponyms, we extracted posts that contain twelve common Japanese nouns and six Japanese toponyms that refer to places outside Japan to construct reference datasets. We refer to the samples for individual keywords (toponyms and nouns) as \emph{keyword subsamples}. In the following, we only use the information about the number of posts tagged inside each grid cell for each sample, and disregard the content of the text or user information. Refer to Table~\ref{tab:variables} for the notation and definition of variables used in this work.

\begin{table}[t]
    \centering
    \caption{\textbf{Variables used in this work.} Subscript $c$ for the grid cell index may be omitted when it is clear.}
    \begin{tabular}{cl}
        \hline
         \textbf{Symbol} & \textbf{Definition}\\
         \hline
         $w$ & Keyword on which the dataset is subsampled\\
         $c$ & Index of grid cell \\ 
         $n_{\mathrm{all}, c}$ & Number of all posts tagged to grid cell $c$ \\
         $n_{w, c}$ & Number of posts containing $w$ and tagged to grid cell $c$ \\
         $N_w$ & Total number of geotagged posts containing keyword $w$; $N_w = \sum_c n_{w, c}$\\
         $A_c$ & Area of grid cell $c$\\
         $\sigma_{\mathrm{all}, c}$ & Density of all posts tagged to grid cell $c$; $\sigma_{\mathrm{all}, c} := n_{\mathrm{all}, c} / A_c$\\
         $\sigma_{\mathrm{res}, c}$ & Resident population density in grid cell $c$ \\
         $\sigma_{\mathrm{emp}, c}$ & Density of employed population (number of personnel working at business sites) in grid cell $c$ \\
         $\sigma_{w, c}$ & Density of posts containing keyword $w$ and tagged to grid $c$; $\sigma_{w, c} := n_{w, c} / A_c$\\ 
         $\phi_{w, c}$ & Occurrence ratio of keyword $w$ among all posts tagged to grid cell $c$; $\phi_{w, c} := n_{w, c} / n_{\mathrm{all}, c}$ \\
         $p_{w, c}$ & (Latent) probability that keyword $w$ is contained in a post tagged to grid cell $c$ \\
         \hline
    \end{tabular}
    \label{tab:variables}
\end{table}

\section{Results}
\subsection*{Spatial distribution of geotagged posts}
\begin{figure}[tb]
    \centering
    \includegraphics[width=\textwidth]{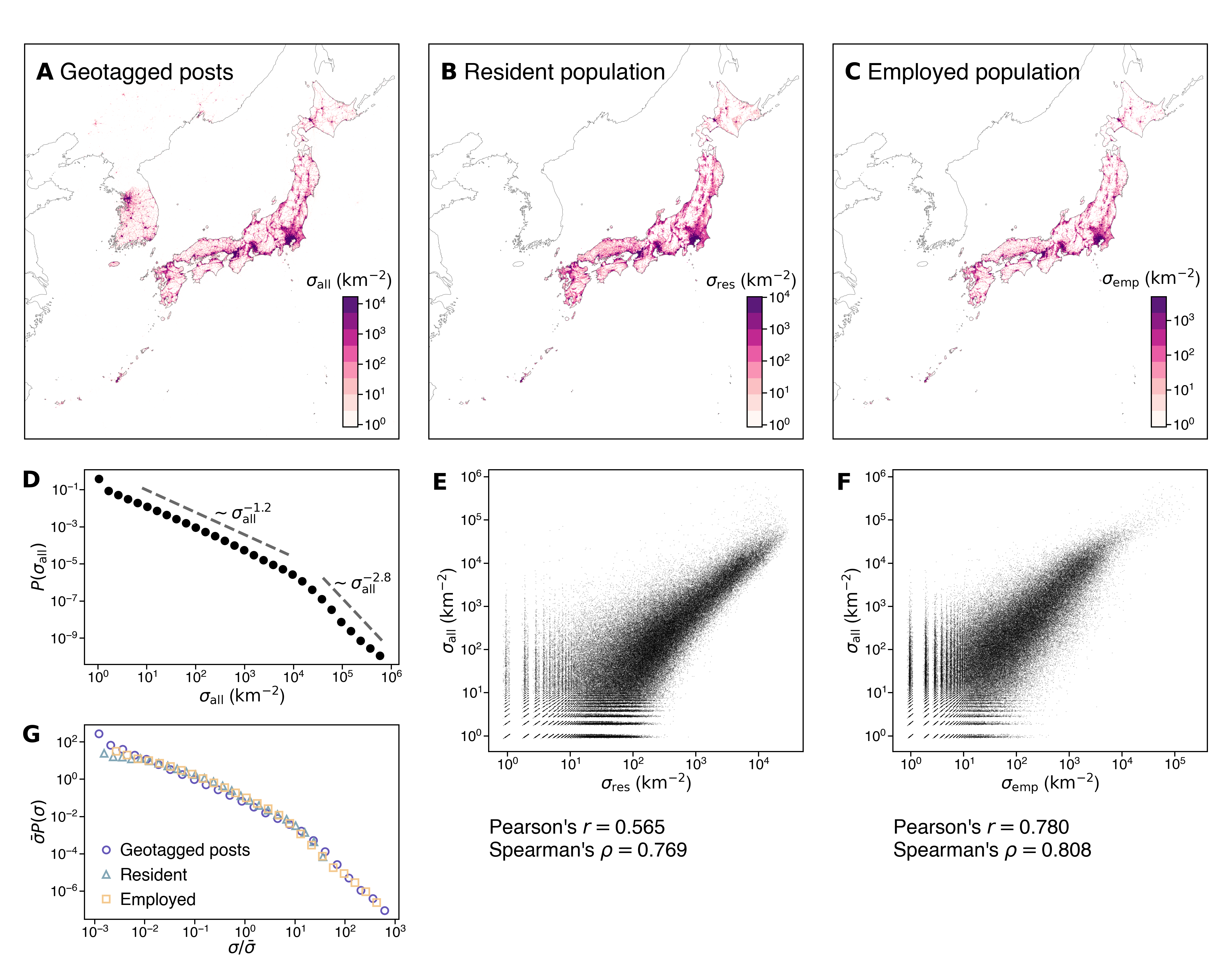}
    \caption{\textbf{The spatial densities of geotagged posts, resident population, and employed population.} \textbf{(A--C)} Geographic distribution of the three densities. Note that the population data are geographically limited inside Japan, while the geotagged posts are sampled in the bounding box of Japan, which also includes neighboring countries. Maps made with Natural Earth (\url{https://www.naturalearthdata.com/}). \textbf{(D}) Probability distribution of density (the number of geotagged posts per unit area). \textbf{(E, F)} Scatter plots showing the correlation between each of the population densities and the geotagged post density. Pearson and the Spearman correlation coefficients are shown below the plot. \textbf{(G)} Probability distributions of the three densities, each rescaled by its mean.}
    \label{fig:density_on_map_all}
\end{figure}

Let us first study the geospatial distribution of the full sample of geotagged posts before looking at the distribution of each toponym subsample (Fig~\ref{fig:density_on_map_all}). The geolocations to which the sampled posts are tagged are not uniformly distributed within the observed area, as shown in Fig~\ref{fig:density_on_map_all}A. A large number of posts are concentrated in relatively few grid cells in large metropolitan areas, such as the city centers of Tokyo and Osaka, while few posts are found in most of the grid cells. The heterogeneity of the spatial distribution of geotagged posts is also evident from the heavy-tailed probability distribution of the number of posts sampled in each grid cell (Fig~\ref{fig:density_on_map_all}D). Namely, it is characterized by two power laws with different exponents: the bulk part is characterized by an exponent of approximately \num{-1.2} while the tail part follows a steeper power law with an exponent of about \num{-2.8}. 

To investigate the origin of this heterogeneous distribution, we compare the geotagged post statistics with census data for the resident population in 2015~\cite{PopCensus} and the employed population in 2016~\cite{EconCensus}. The employed population is defined as the number of permanent or temporary employees, self-employed individuals, contractors, and unpaid staff in family-owned businesses, whose main working site is located in each grid cell. From Figs.~\ref{fig:density_on_map_all}B and ~\ref{fig:density_on_map_all}C, one can see the similarity in the spatial distribution between the densities of geotagged posts, resident population, and employed population. In fact, the geotagged post density is strongly correlated with the population densities (Figs.~\ref{fig:density_on_map_all}E, ~\ref{fig:density_on_map_all}F). The correlation with the post density is stronger for the employee population density than for the resident population density, both in terms of Pearson's correlation and Spearman's rank correlation. This may suggest that social media posts are made more in the daytime than at night.  Figure~\ref{fig:density_on_map_all}G shows the probability density functions of the geotagged post density, the resident population density, and the employee population density collapse to one another when they are scaled by the average density, strongly suggesting that the uneven distribution of population gives rise to the heterogeneity in the spatial distribution of geotagged posts.

\subsection{Spatial distributions of toponym subsamples}
We now turn to our main question: How are geotagged posts containing each toponym, i.e., each \emph{toponym subsample}, spatially distributed? To address this question, we show the results for \emph{Fukuoka}, a toponym referring to the sixth largest city in Japan as of 2019, located in the southwestern part of the country, as an illustrative example. 

\begin{figure}[tb]
    \centering
    \includegraphics[width=0.83\textwidth]{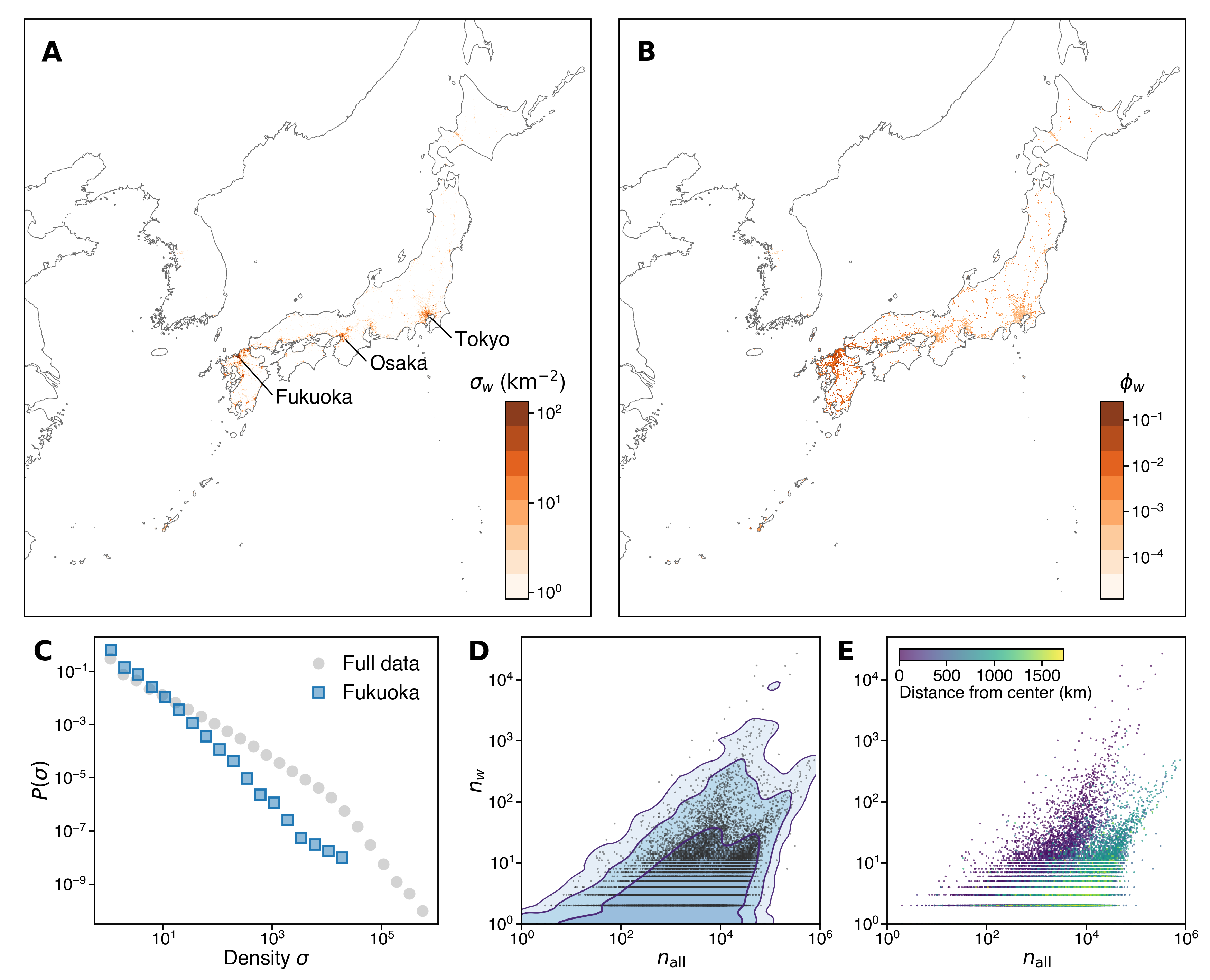}
    \caption{\textbf{Occurrence pattern of \emph{Fukuoka}.} \textbf{(A)} Geographic distribution of density $\sigma_w$. \textbf{(B)} Geographic distribution of occurrence ratio $\phi_w$. \textbf{(C)} Probability distributions of spatial density for all geotagged posts (in gray) and for \emph{Fukuoka} (in blue). \textbf{(D)} Scatter plot showing the relationship between the total number of geotagged posts $n_\mathrm{all}$ and the number of posts containing \emph{Fukuoka} $n_w$ in each grid cell. The lines represent contours along which the density of points (kernel density estimate) on the double logarithmic scale is constant. The region inside each contour, from dark to light colors, contains 90.0\%, 99.0\%, and 99.9\% of the data points, respectively. \textbf{(E)} The same scatter plot as panel D, but with points colored by the distance between the grid cell and the center $O_w$. Maps made with Natural Earth (\url{https://www.naturalearthdata.com/}).}
    \label{fig:fukuoka}
\end{figure}

Figure~\ref{fig:fukuoka}A shows the spatial distribution of the occurrence of \emph{Fukuoka}, while Fig~\ref{fig:fukuoka}C shows the probability density functions of the gridwise occurrence density $\sigma_w$. Both figures suggest that the \emph{Fukuoka} occurs heterogeneously across different grid cells. The same observation can be made for other toponyms, as shown in Fig~\ref{fig:density_pdfs} in Supporting Information. It is noteworthy that toponym subsamples of different sizes follow the same scaling, which implies that the spatial heterogeneity does not depend on the popularity of toponyms. On the other hand, each distribution is characterized by a single power law, which is a clear deviation from the scaling observed for the full sample. 

While the spatial density (occurrence per unit area) of posts containing the word \emph{Fukuoka} is high in Fukuoka and neighboring areas, it is also high in other large metropolitan areas such as Tokyo and Osaka, which are geographically distant from Fukuoka. However, this is presumably just by chance due to the large total number of posts tagged in these areas. To account for the difference in total sample size $n_\mathrm{all}$ in each grid cell, we normalize the toponym subsample size $n_w$ by $n_\mathrm{all}$ and obtain the fraction of occurrence of the toponym. We clearly see that posts contain the word \emph{Fukuoka} with higher probabilities in the region around the city of Fukuoka (Fig~\ref{fig:fukuoka}B).

To further understand the relationship between the full sample and the toponym subsamples, we create a scatter plot for each toponym as in Fig~\ref{fig:fukuoka}D, where the horizontal axis represents $n_\mathrm{all}$ and the vertical axis corresponds to $n_w$. Apart from the general trend that $n_w$ increases with $n_\mathrm{all}$, we see that there are two distinct branches of scaling, with one increasing faster than the other. That is, $n_w$ increases as a function of $n_\mathrm{all}$ in two different ways. This two-branch scaling behavior is not specific to \emph{Fukuoka} but is widely seen for different toponyms across various degrees of popularity and granularity (Fig~\ref{fig:all_topo_many_scatter} in Supporting Information). In contrast, keyword subsamples for common nouns such as \emph{wallet} and toponyms that refer to places outside the observed area such as \emph{Hawaii} do not exhibit heterogeneity in spatial distribution, and the relationship between $n_\mathrm{all}$ and $n_w$ only shows a single scaling (see Figs.~\ref{fig:all_toponym_example}B and  \ref{fig:all_toponym_example}D; see also \ref{fig:all_topo_many_scatter} for the scatter plots for all keywords sampled in this study). This suggests that the two-branch scaling is unique to the toponyms that refer to places within the observed area and is not present in other types of words. Naturally, we expect that such a pattern stems from the geospatial specificity of toponym use.

In particular, we hypothesize that one of the two scaling branches observed for toponym subsamples corresponds to the distribution of the toponym that is spatially specific to the area it refers to, while the other branch represents the use of the toponym that is not spatially specific, similar to that of common nouns. To test this hypothesis, we first need to identify the area that the toponym refers to. Instead of relying on external sources such as gazetteers, we do this in a data-driven way: we define the \emph{center} of the geographic distribution of toponym $w$ as the grid cell with the highest frequency of $w$:
\begin{equation}
    O_w = \operatorname*{argmax}_{c \in S_w}\, n_{w, c}, 
\end{equation}
where $S_w = \{c \mid \phi_{w, c} \geq \epsilon\}$ is the set of cells with an occurrence ratio equal to or greater than $\epsilon$. This condition prevents a cell from being selected as the center merely due to the large total number of posts. Here we set $\epsilon=0.01$, i.e., toponym $w$ must appear in at least 1\% of all posts tagged to a cell for the cell to be included in $S_w$. As we can see in Fig~\ref{fig:fukuoka}E, the grid cells that constitute the two branches are clearly distinct in terms of distance from the center of the toponym subsample. The branch with faster scaling is geographically closer to the center while the branch with slower scaling is relatively far from the center. This corroborates our hypothesis that the two-branch scaling arises from the spatial specificity of the toponym.

\subsection{Location-independent model}
The results in the previous section give us an intuitive understanding of the geographic distribution of toponyms. In the next two sections, we present a model-based analysis to validate our intuition and to characterize the empirical observation in a quantitative way. 
Specifically, we introduce a model where each post tagged to location $c$ contains word $w$ with probability $p_{w, c}$ that may depend on $c$. We assume that the occurrence of word $w$ in each post is an independent event, i.e., a Bernoulli trial.
The number $n_{w, c}$ of posts in grid cell $c$ that contains word $w$ out of $n_{\mathrm{all},c}$ total posts follows a binomial distribution:
\begin{equation}
    P(n_{w, c} \mid n_{\mathrm{all}, c}, \, p_{w, c}) = \binom{n_{\mathrm{all}, c}}{n_{w, c}} p_{w, c}^{n_{w, c}} (1 - p_{w, c})^{n_{\mathrm{all}, c} - n_{w, c}}.
    \label{eq:binomial_model}
\end{equation}
This model essentially posits that the underlying mechanism of the occurrence of a word can be summarized by probability $p_{w, c}$ specific to each grid cell $c$. Importantly, it relies on the simplifying assumption that the occurrence of a word in one post is independent of its occurrence in other posts, and that the dependence between the occurrence of different words within the same post can also be neglected.

We first examine if the empirical pattern can be explained by the simplest version of this binomial model which assumes that $p_{w, c} = p_w$ for all grid cell $c$; that is, the toponym occurs with a constant probability independent of location. 
The unbiased and maximum likelihood estimator for $p_w$ can be obtained simply by dividing the size of the toponym subsample by the number of all geotagged posts: $\hat{p}_w = N_w / N_\mathrm{all}$. 

\begin{figure}[tb]
    \centering
    \includegraphics[width=\textwidth]{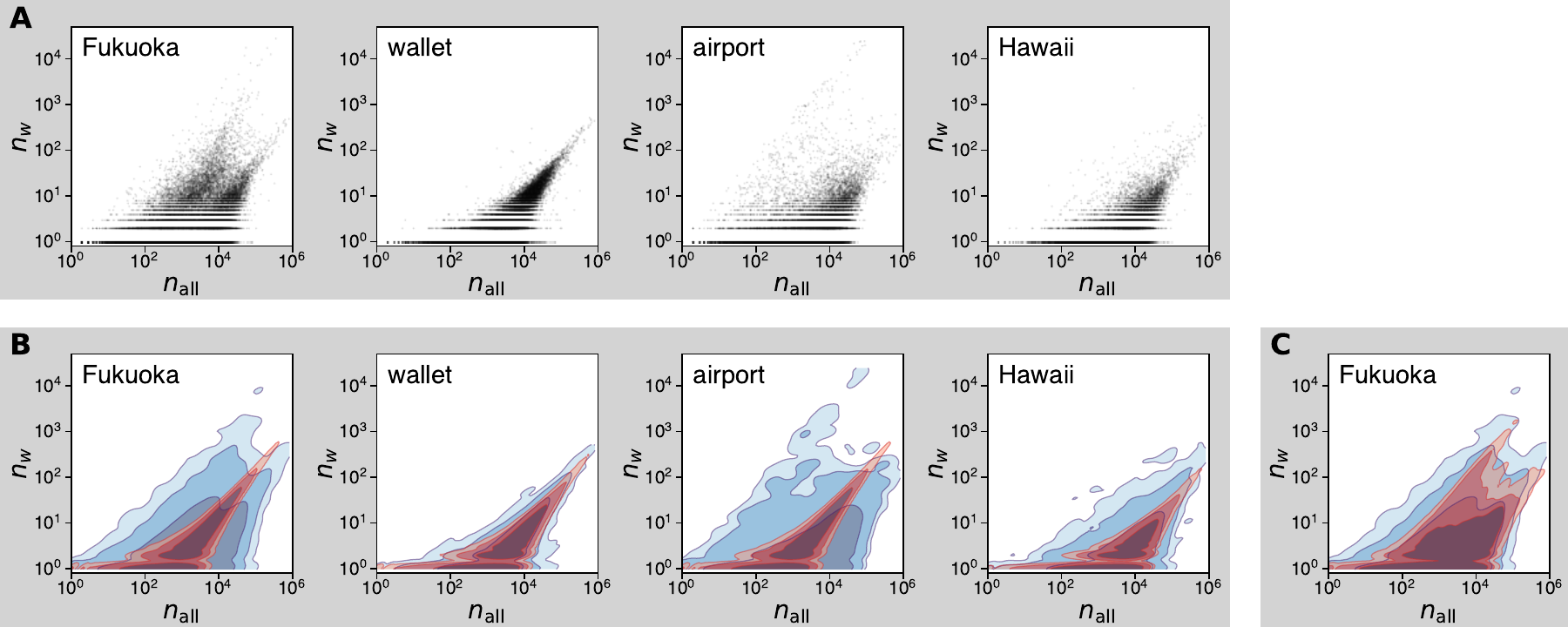}
    \caption{\textbf{Relationship between $n_w$ and $n_\mathrm{all}$ in empirical and model distributions.} \textbf{(A)} Scatter plots for different keywords. \textbf{(B, C)} Kernel density profiles of empirical and model distributions. 
    In each panel, the red contours show the density profile obtained from the location-independent model (B) or the core-periphery model (C), fitted to the empirical occurrence pattern of each word, represented by the blue contours. As in Fig.~\ref{fig:fukuoka}D, the region inside each contour, from dark to light colors, contains 90.0\%, 99.0\%, and 99.9\% of the data points, respectively. Note that the scatter plot of empirical data and its density profile for \emph{Fukuoka} are identical to those in Fig~\ref{fig:fukuoka}D.}
    \label{fig:all_toponym_example}
\end{figure}

Figure~\ref{fig:all_toponym_example}A shows the distribution of occurrence of \emph{Fukuoka} against $n_\mathrm{all}$ in empirical data and the expectation from the location-independent binomial model. We observe that the empirical distribution is not consistent with the model. In particular, the model does not reproduce the two-branch scaling behavior seen in the empirical data and exhibits a single scaling instead. We confirm the same kind of discrepancy for all the domestic toponyms we studied (see Fig~\ref{fig:all_topo_many_lim} in Supporting Information for the full results).

The implication of this observation becomes apparent through juxtapositions against patterns for common nouns and foreign nouns. The location-independent model shows a close agreement with the empirical data for \emph{wallet} (Fig~\ref{fig:all_toponym_example}B), implying that the word indeed occurs at a constant rate in every grid cell. In general, we find that the empirical data and the location-independent model are in fairly good agreement for many common nouns that refer to objects (e.g., \emph{telephone}) or abstract concepts (e.g., \emph{society}). For a comparison between the model and the empirical distributions for these words, see Fig~\ref{fig:all_topo_many_lim} in Supporting Information. On the other hand, there is a large deviation between the empirical and model distributions for the word \emph{airport} (Fig~\ref{fig:all_toponym_example}C), suggesting that the occurrence at a constant rate is not a shared characteristic among all common nouns. 
This can be explained by the fact that common words such as \emph{airport} are often used in combination with toponyms (e.g. \emph{Narita Airport}), and are therefore more likely to be used in specific places.
Finally, for foreign toponyms, such as \emph{Hawaii} shown in Fig~\ref{fig:all_toponym_example}D, we observe the single scaling pattern in both the empirical and model distributions, although the empirical distributions generally show slightly greater variance than the model. While these toponyms are not semantically associated with a specific place in the observed area, they may occur with higher probability around international airports from which people travel to the places these toponyms refer.

Beyond visual inspection, the discrepancy between the location-independent model $P$ and the data can be quantified using relative entropy, also known as the Kullback--Leibler (KL) divergence. Intuitively, relative entropy measures the dissimilarity from one probability distribution to another. Here, we wish to quantify the dissimilarity of the empirical data from the model, each of which is a probability distribution on the total number of posts in each cell and the number of posts with word $w$ in each cell. Specifically, we employ a modified version of relative entropy, denoted by $D_\mathrm{KL} (\tilde{Q}_w \parallel P_w)$, that takes into account the cells with at least one occurrence of $w$. We elaborate on this particular definition and provide the rationale for our choice in Supporting Information.

\begin{figure}[t]
    \centering
    \includegraphics[width=\textwidth]{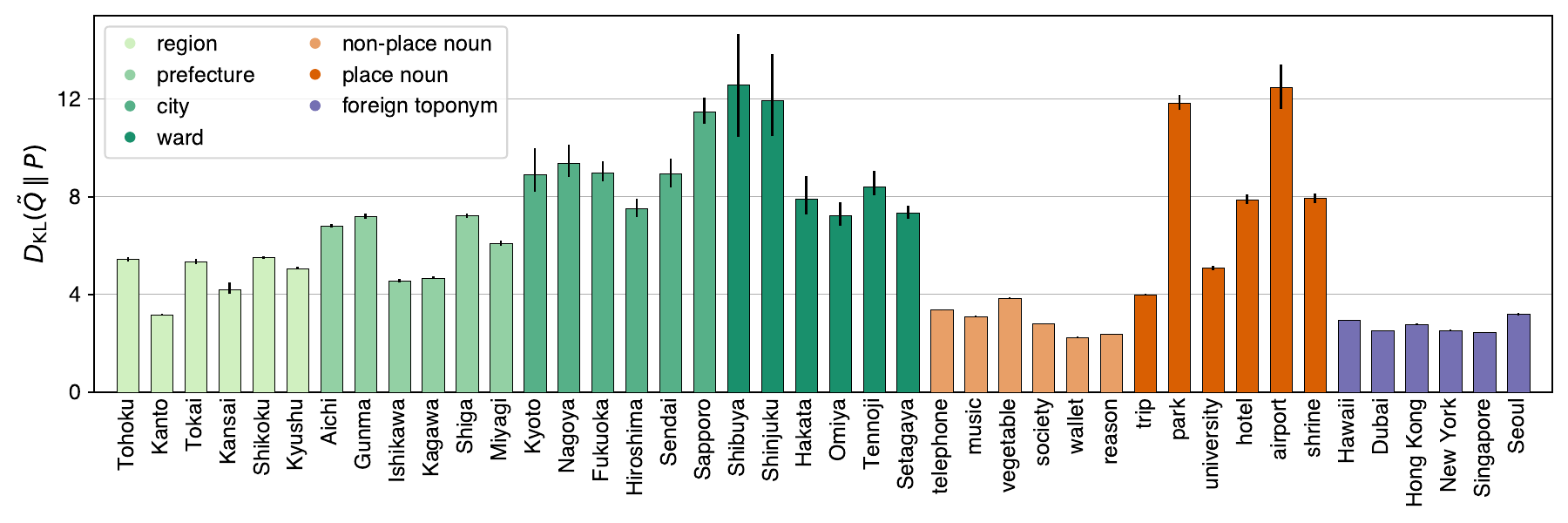}
    \caption{\textbf{Dissimilarity of empirical data from the location-independent model.} The dissimilarity is evaluated by relative entropy $D_\mathrm{KL} (\tilde{Q}_w \parallel P_w)$. For each word, 3000 grid cells are randomly sampled 50 times. The error bar shows the 95\% confidence interval.}
    \label{fig:nonzero_relative_entropy}
\end{figure}

As shown in Fig~\ref{fig:nonzero_relative_entropy}, relative entropy $D_\mathrm{KL} (\tilde{Q}_w \parallel P_w)$ clearly differentiates domestic toponyms and place nouns from foreign toponyms and common nouns without place connotations. All domestic toponyms except \emph{Kanto} and all place nouns (\emph{park}, \emph{university}, \emph{hotel}, \emph{airport}, and \emph{shrine}) are characterized by relatively large values of relative entropy, namely $D_\mathrm{KL} (\tilde{Q}_w \parallel P_w) > 4$, implying that the empirical distribution is highly dissimilar from the location-independent binomial model for these words. Common nouns such as \emph{trip} and \emph{vegetable} are on the borderline ($D_\mathrm{KL} (\tilde{Q}_w \parallel P_w) \simeq 4$); relative entropy for other common nouns and foreign toponyms are significantly smaller. 

The comparison with the location-independent model reveals how sensitive and specific the occurrences of toponyms and common nouns are to geographic locations. For toponyms, the spatial specificity may be readily observable by visualizing the geographic distribution of occurrence ratio $\phi_w$, as in Fig~\ref{fig:fukuoka}B. However, identifying location-specific common nouns may be more subtle, as these nouns are usually associated not with a single place but with multiple places across the observed area. As a result, their geographic distributions may appear visually indistinguishable from those of non-place nouns.
In such cases, quantifying the dissimilarity from the binomial model through $D_\mathrm{KL} (\tilde{Q}_w \parallel P_w)$ can serve as a good indicator of the geospatial specificity of word occurrence.

\subsection{Core-periphery model}
So far, we have shown that the location-independent binomial model cannot reproduce the large variance and two-branch scaling behavior seen in the empirical distribution of domestic toponyms. Let us now take a step toward realism and discuss a more flexible modeling framework to account for the empirically observed geographic distribution of toponyms. We consider the \emph{location-dependent} binomial model, that is, we assume that the occurrence probability $p_{w, c}$ in Eq~\eqref{eq:binomial_model} can vary for different grid cells. 

A straightforward implementation of the location-dependent model would be to allow the occurrence probabilities for different cells to be independent of each other. In this case, the value of $p_{w, c}$ can be estimated individually for each $c$ and its unbiased maximum likelihood estimator would be equal to the occurrence ratio $\phi_{w, c} = n_{w, c} / n_{\mathrm{all}, c}$. However, since this model has as many parameters as the number of grid cells, it provides little insight into the spatial patterns of toponym occurrence. What would be more interesting to us is a model that captures the empirical observation with a smaller number of parameters. 

\begin{figure}[tb]
    \centering
    \includegraphics[width=0.98\textwidth]{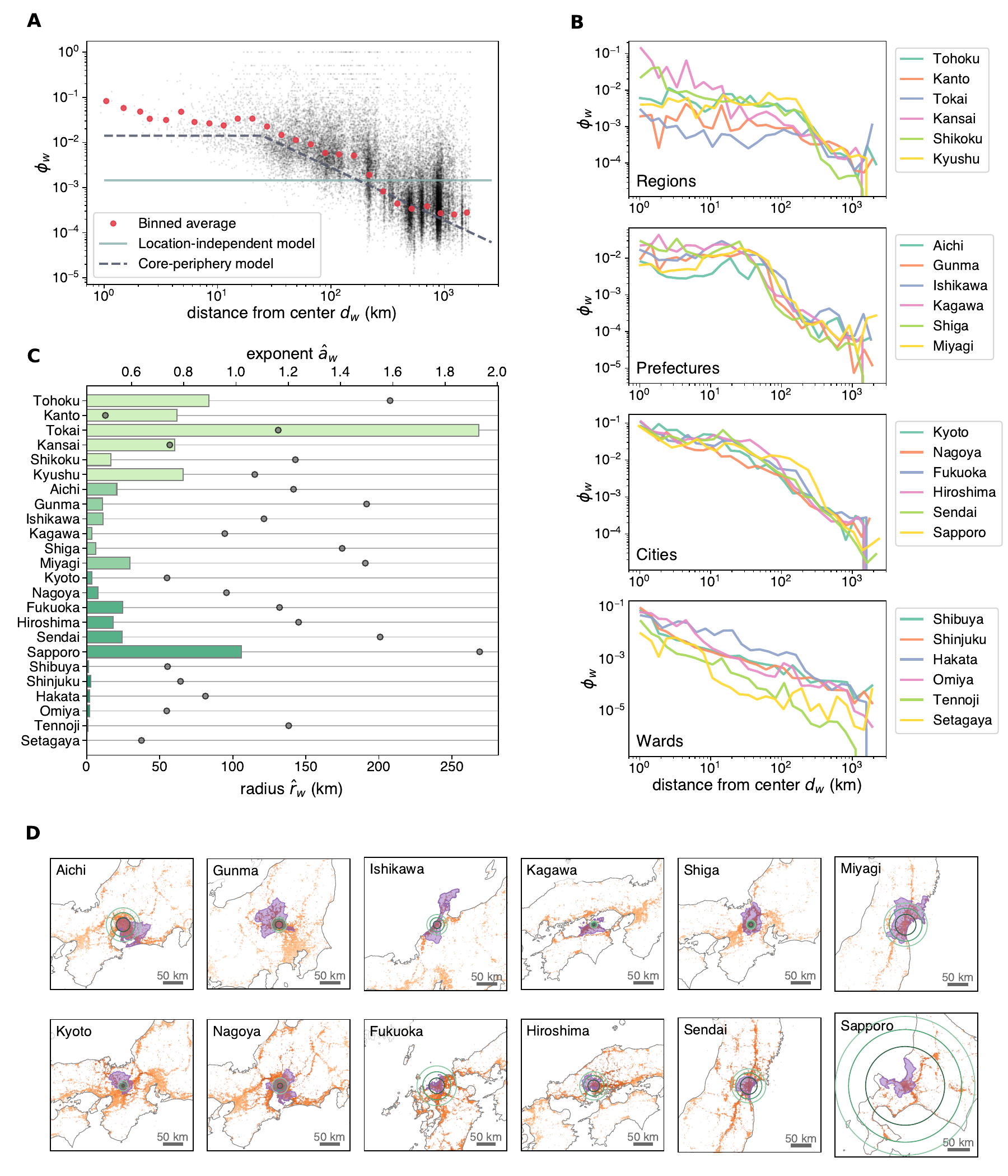}
    \caption{\textbf{Core-periphery patterns of toponym occurrence.} \textbf{(A)} Occurrence ratio $\phi_w$ of \emph{Fukuoka} against distance $d_w$ from center $O_w$ (small black dots), overlaid with the average for each logarithmic bin (red circles). The solid and dashed lines represent the maximum likelihood fits of the location-independent and core-periphery binomial models. \textbf{(B)} Average occurrence ratio as a function of $d_w$ for all the domestic toponyms studied in this work. \textbf{(C)} Maximum likelihood estimator of the core-periphery model parameters for each toponym. We represent estimated radius $\hat{r}_w$ by bars colored according to the category of the toponym (lower axis) and estimated exponent $\hat{a}_w$ by gray circles (upper axis). The standard errors are omitted as they are too small to be meaningfully visualized. \textbf{(D)} The fitted core-periphery model compared to the administrative/metropolitan area. The innermost circles in dark green represent the core boundary (distance $r_w$ from the center $O_w$) and the two outer circles in lighter green denote the distance at which the occurrence probability $p_{w, c}$ is equal to one half and one third of the probability in the core $q_w$, respectively. The areas shaded in purple indicate the administrative area of each prefecture (top row) and the metropolitan employment area of each city (bottom row). Maps made with Natural Earth (\url{https://www.naturalearthdata.com/}).}
    \label{fig:core_periphery}
\end{figure}

Let us recall that the results in Fig~\ref{fig:fukuoka} suggest that a post in grid cell $c$ is more likely to contain a toponym $w$ if $c$ is within the area that $w$ refers to. Indeed, the occurrence ratio $\phi_{w, c}$ plotted against the geodesic distance $d_{w, c}$ between the center $O_w$ and cell $c$ shows an overall decreasing trend (Fig~\ref{fig:core_periphery}A). The binned averages of $\phi_{w, c}$ imply that the occurrence probability decays as a power-law function of distance from the center, especially at long distances. Moreover, by grouping toponyms according to their administrative level as shown in Fig~\ref{fig:core_periphery}B, one can see that the onsets of the power law differ according to the granularity of each toponym. For toponyms referring to higher-level units, such as regions and prefectures, which are typically larger in area, the power-law decay starts at larger distances while the average occurrence ratios are relatively stable at smaller distances. Conversely, toponyms denoting small administrative units, such as wards, exhibit power-law behavior starting at short distances with no noticeable plateau regime.

Motivated by these observations, we propose a simple location-dependent variant of the binomial model (Eq~\eqref{eq:binomial_model}) that assumes the presence of a \emph{core}, the area in which the occurrence probability $p_{w, c}$ is high, and a \emph{periphery}, grid cells that are geographically distant from the center and characterized by lower $p_{w, c}$. Namely, the occurrence probability is assumed to be constant within a certain distance from the center and to decrease as a power law outside of this range:
\begin{equation}
    p_{w, c} = 
    \begin{cases}
        q_w & \text{for } d_{w, c} \leq r_w, \\
        q_w \left[\dfrac{d_{w, c}}{r_w}\right]^{-a_w} & \text {for } d_{w, c} > r_w.
    \end{cases}
\end{equation}
This model has three free parameters: $q_w$ is the occurrence probability in cells within a radius of $r_w$ from the center, and $a_w$ denotes the exponent of the power-law decay outside the range. In Fig~\ref{fig:core_periphery}A, we visualize the behavior of this model as a function of distance from the center in comparison to that of the location-independent model.

For each toponym $w$, the parameter values of the core-periphery model are estimated by maximizing the log-likelihood function 
\[
\mathcal{L}(q_w, r_w, a_w) = \sum_c \log P(n_{w,c} \mid n_{\mathrm{all}, c}, \, p_{w, c}).
\]
The numerical optimization is carried out using SciPy's \verb|scipy.optimize.minimize| function with the Nelder-Mead algorithm as the solver. To reduce the risk of the solution being trapped in a local minimum, the parameter $r_w$ is initialized with four distinct values: \SI{10}{km}, \SI{20}{km}, \SI{40}{km}, and \SI{80}{km}.

\begin{figure}[tb]
    \centering
    \includegraphics[width=0.6\textwidth]{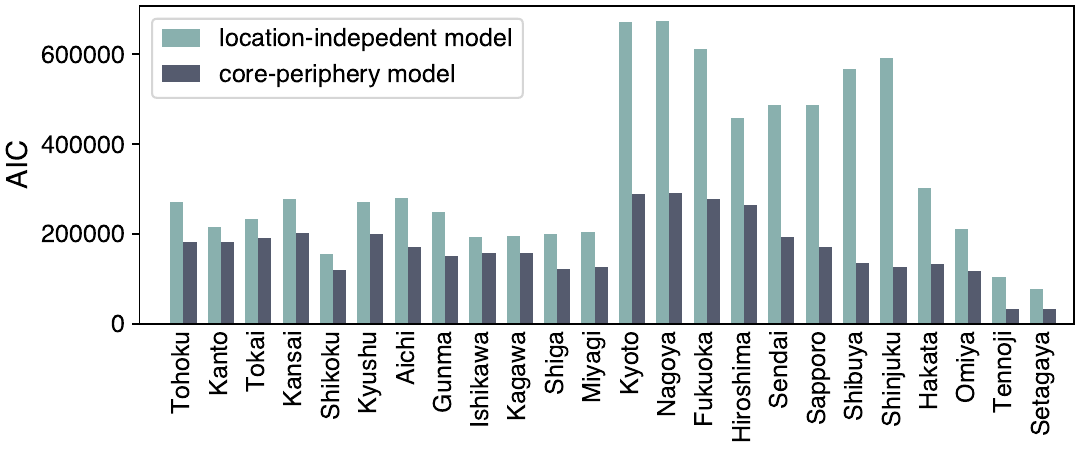}
    \caption{\textbf{Goodness of fit evaluated by Akaike information criterion (AIC).}}
    \label{fig:aic}
\end{figure}

The fitted model shows, in general, a better agreement to the data in terms of the relationship between $n_\mathrm{all}$ and $n_w$, as shown in Fig~\ref{fig:all_toponym_example}B. For the results for all domestic toponyms studied in this work, see Fig~\ref{fig:all_topo_many_cpm} in Supporting Information.
In particular, this model reproduces the two-branch scaling behavior of the empirical data. This significant improvement from the location-independent model is also evidenced quantitatively by the decrease in the Akaike Information Criterion (AIC) for all the toponyms studied in this work (Fig~\ref{fig:aic}).

The advantage of the core-periphery model is that it conforms to an intuitive interpretation: the radius $r_w$ can be seen as the extent of the area to which the toponym refers, i.e. the core. Outside this core area, users refer to the place less often, but the decrease in probability is gradual as a function of distance and slow enough to be modeled as a power law (compared to, e.g., an exponential decay). To verify this interpretation, we compare the estimated core (the area within the estimated radius $\hat{r}_w$ from the center $O_w$) of each toponym with the extent of the administrative unit or metropolitan area to which it refers (Fig~\ref{fig:core_periphery}D). For all toponyms, the model identifies the center within the area to which each toponym refers, although the sizes of the cores vary and do not necessarily coincide with the administrative boundaries. For many toponyms, the core area detected by the model is smaller than the administrative area. This may indicate that users are more likely to make geotagged posts with these toponyms when they are in the central city of a prefecture or in the central area of a city. For \emph{Fukuoka} and \emph{Sendai}, the core has a geographic scale similar to the metropolitan area, suggesting that the use of these two toponyms is aligned with the extent of the corresponding metropolitan area. In the case of \emph{Sapporo}, the core is larger than the metropolitan area. This could indicate that users tend to associate the toponym with a larger area; however, it is also possible that this is because the estimated parameter represents one of the many local maxima in the likelihood landscape.

In Fig~\ref{fig:core_periphery}C, we show the estimated values of the radius and exponent for each of the 24 domestic toponyms we study. The radii of the toponyms vary according to the size of the area they denote. Region names are associated with larger cores, typically ranging from \SI{50}{km} to \SI{100}{km} in radius, which is consistent with the spatial scale of regions. In contrast, ward names are characterized by much smaller cores, with radii less than \SI{5}{km}. Prefectures and cities fall in between, with core radii typically between \SI{10}{km} and \SI{30}{km}. The value of the exponent of the decay outside the core also varies from one toponym to another; however, it does not seem to correlate clearly with other quantities, such as the frequency of toponym occurrence. 

\section{Discussion and Conclusions}
In this article, we investigated the geographic patterns of toponym occurrence in social media using a dataset of geotagged Twitter posts. We found that the heterogeneous geographic distribution of geotagged posts is highly correlated with the population, especially with the employed population. This implies that the geotagged posts are, on the whole, representative of the language use in the population. The occurrence of each toponym in these geotagged posts is also characterized by geographic heterogeneity. Moreover, we found that the relationship between the total number of posts and the number of posts containing toponyms shows a distinctive scaling pattern. Comparison with patterns for common nouns and foreign toponyms suggests that this scaling pattern originates from the spatial specificity of toponym occurrence, which is successfully quantified by the dissimilarity from the location-independent model. 

Finally, we presented the core-periphery model, which assumes a location-dependent occurrence probability of toponyms. Despite its simplicity with only three fitting parameters, this model can reasonably reproduce the empirically observed geographic distributions of toponym occurrence. This implies the following: First, each toponym has a core, i.e., a geographic area in which the toponym occurs with the highest probability, which can be regarded as the area that users collectively identify with the toponym. This interpretation is supported by the fact that the core is larger for the names of regions than for the names of cities and wards. Second, outside this core, the occurrence probability decreases slowly with distance following a power law. 

Our findings may indicate that human attention, cognition, and representation of geographic space respond nonlinearly to distance~\cite{Montello2009}. It is reminiscent of Tobler's first law of geography: ``everything is related to everything else, but near things are more related than distant things''~\cite{Tobler1970}. In this context, our findings can be seen as another example of the distance decay phenomenon, which has been observed in various aspects of human behavior such as commuting~\cite{Iacono2008, Helminen2012, Halas2014}, tourism~\cite{McKercher2008, Hooper2015, McKercher2018}, and crime~\cite{Rengert1999, Kent2006, Townsley2010}. The concept of distance decay, and its more sophisticated formulation, the gravity model, has also been used in archaeology and history~\cite{Tobler1971, Renfrew1977}, attesting to its universal applicability in describing human activities. Inspired by developments in geography, distance decay and gravity models have been adopted in linguistics to explain variations in pronunciation between different dialects~\cite{Trudgill1974, Nerbonne2007} and language evolution via lexical replacement~\cite{Cavalli-Sforza1986}. However, unlike other language elements that are generally geography-neutral, each toponym is intrinsically associated with a specific geographic point or area. As such, our results present a unique variant of Tobler's first law in language, one that cannot be characterized by autocorrelation or other similarity measures~\cite{Miller2004, Waters2018}.

We note that our modeling approach does not aim to precisely replicate empirical observations of toponym occurrence. Rather, our models serve as a reference against which empirical data can be compared. As such, they simplify some aspects of the real-world toponym occurrence patterns. For example, in the core-periphery model, the model is isotropic, that is, the core has a circular boundary and the occurrence probability decreases uniformly in all directions. In reality, however, the area denoted by the toponym is not necessarily circular and may have elongated or irregular shapes, and the decay outside the core may not be isotropic due to geographical or transportation constraints. We also assumed no correlations between the occurrence of different toponyms, although there may be competitive or synergistic interactions between them that influence their occurrence patterns. It is because of these simplifications that our approach is able to provide an interpretable framework that allows us to identify the essential elements of the geospatial toponym distributions. 

We also remark on the possibility that geotagged posts on Twitter may not be an unbiased, representative sample of the language use of the general population. This issue can be divided into two questions: whether the geotagged Twitter users can be considered a good proxy for the population at large~\cite{Hecht2014, Pavalanathan2015, Malik2015, Anselin2016}, and whether the language use in geotagged Twitter posts is consistent with language use in other contexts~\cite{Li2013, Wartmann2018}. The effect of the first problem on our results is presumably relatively small compared to other work using geotagged Twitter posts to study language use, for two reasons: (i) we focus only on the content of the posts without correlating them with user demographics, and (ii) the age and socioeconomic status of users are unlikely to significantly affect their use of toponyms. It is however possible that urban toponyms are overrepresented due to population biases, which could affect the geographic distribution of toponyms. In this work, we focused on relatively large cities and wards within them, but whether our findings generalize to the names of smaller towns and villages needs to be carefully examined in future research. The second question concerns the generalizability of our findings to the use of toponyms in other contexts. Indeed, prior research has shown that, for certain linguistic features, such as emoji and hashtags, the same user may exhibit different styles on different social platforms~\cite{Marko2022}. However, we expect that toponyms are unlikely to be strongly influenced by such contextual differences. Toponyms are a linguistic element that is relatively stable and resistant to change~\cite{PeronoCacciafoco2023}, and it is reasonable to assume that this stability stems from the low synchronic variability of toponym use in the population.

Lastly, we note that our models are phenomenological and do not account for the microscopic underpinnings behind the observation, such as the behavior of individual users or posts. Moreover, gaining a comprehensive understanding of the use of toponyms---how they reflect the interaction between people and environment, how they shape and reinforce people's identity, and how they are affected by urban planning and place branding---would require historical, ecological, cultural, and economic analyses~\cite{Conedera2007, Radding2010, Rose-Redwood2010, Hakala2015, Light2015, Capra2016, Atik2017, Rose-Redwood2019}. These aspects are abstracted away in the present study, where our focus is to establish general empirical laws that govern the spatial distribution of toponyms. The simplified models we propose are designed to serve the purpose of quantitative, so-called \emph{extensive} analysis~\cite{Tent2015}. Future work should aim to integrate these qualitative factors with our quantitative framework to investigate the dynamics underlying toponym usage and distribution.

\begin{acknowledgments}
T.H. acknowledges the computational resources provided by the Aalto Science-IT project.
N.F. was supported by JSPS KAKENHI Grant Number 24K03007 and JST PRESTO Grant Number JPMJPR21RA, Japan.

\noindent \textbf{Data availability:} All data and code necessary to reproduce the findings of this paper are deposited and publicly available at \url{https://doi.org/10.5281/zenodo.13860968} and \url{https://github.com/takayukihir/geotagged-tweets}. Due to the Twitter's terms of service, we are unable to redistribute the raw text of the posts we have obtained from the Twitter API. Instead, the data on the number of geotagged posts aggregated at the level of basic grid square cells are available in the above repository. The resident and employed population data are based on 2015 Population Census and 2016 Economic Census for Business Activity, respectively, and are available from the Statistics Bureau of Japan (\url{https://www.stat.go.jp/english/data/kokusei/2015/summary.html}, \url{https://www.stat.go.jp/english/data/e-census/2016/outline.html}).
\end{acknowledgments}

\clearpage

\begin{appendix}
\renewcommand{\thefigure}{SA\arabic{figure}}
\setcounter{figure}{0}
\renewcommand{\theequation}{S\arabic{equation}}
\setcounter{equation}{0}
\renewcommand{\thesubsection}{S\arabic{subsection}}

\section*{Supporting information}

\section{Definition of relative entropy}

In the main text, we use relative entropy (Kullback--Leibler (KL) divergence) to measure the dissimilarity between the empirical and model distributions. Here, we describe the definition of relative entropy in detail. 

Each distribution is defined in the two-dimensional space of the total number of posts per cell, denoted by $n_\mathrm{all}$, and the number of posts with word $w$, denoted by $n_w$. Specifically, we only consider the values of $n_\mathrm{all}$ that are empirically observed, and the values of $n_w$ equal to or smaller than $n_\mathrm{all}$. That is, the support of each distribution is given by 
\[
\{(n_\mathrm{all}, n_w) \in \mathbb{N} \times \mathbb{N} \mid (\exists c \in \Gamma) [n_\mathrm{all}=n_{\mathrm{all}, c}], n_w \leq n_\mathrm{all}\}
\] 
for a set of grid cells $\Gamma$. Given a set of empirical numbers of posts $\{n_{w, c}\}$ containing word $w$ for $c \in \Gamma$, relative entropy is defined as
\begin{equation}
    D_\mathrm{KL} (Q_w \parallel P_w) = \frac{1}{|\Gamma|}\sum_{c \in \Gamma} 
    \left[\log Q_w (n_{w, c} \mid n_{\mathrm{all}, c}) - \log P(n_{w, c} \mid n_{\mathrm{all}, c}, \, \hat{p}_{w})\right], 
\end{equation}
where $P_w := P(n_w \mid n_\mathrm{all}, \hat{p}_{w})$, and $Q_w (n_w \mid n_\mathrm{all})$ denotes the empirical distribution of $n_w$ for given $n_{\mathrm{all}}$ over the set of grid cells $\Gamma$. 

\begin{figure}[tbh]
    \centering
    \includegraphics[width=\textwidth]{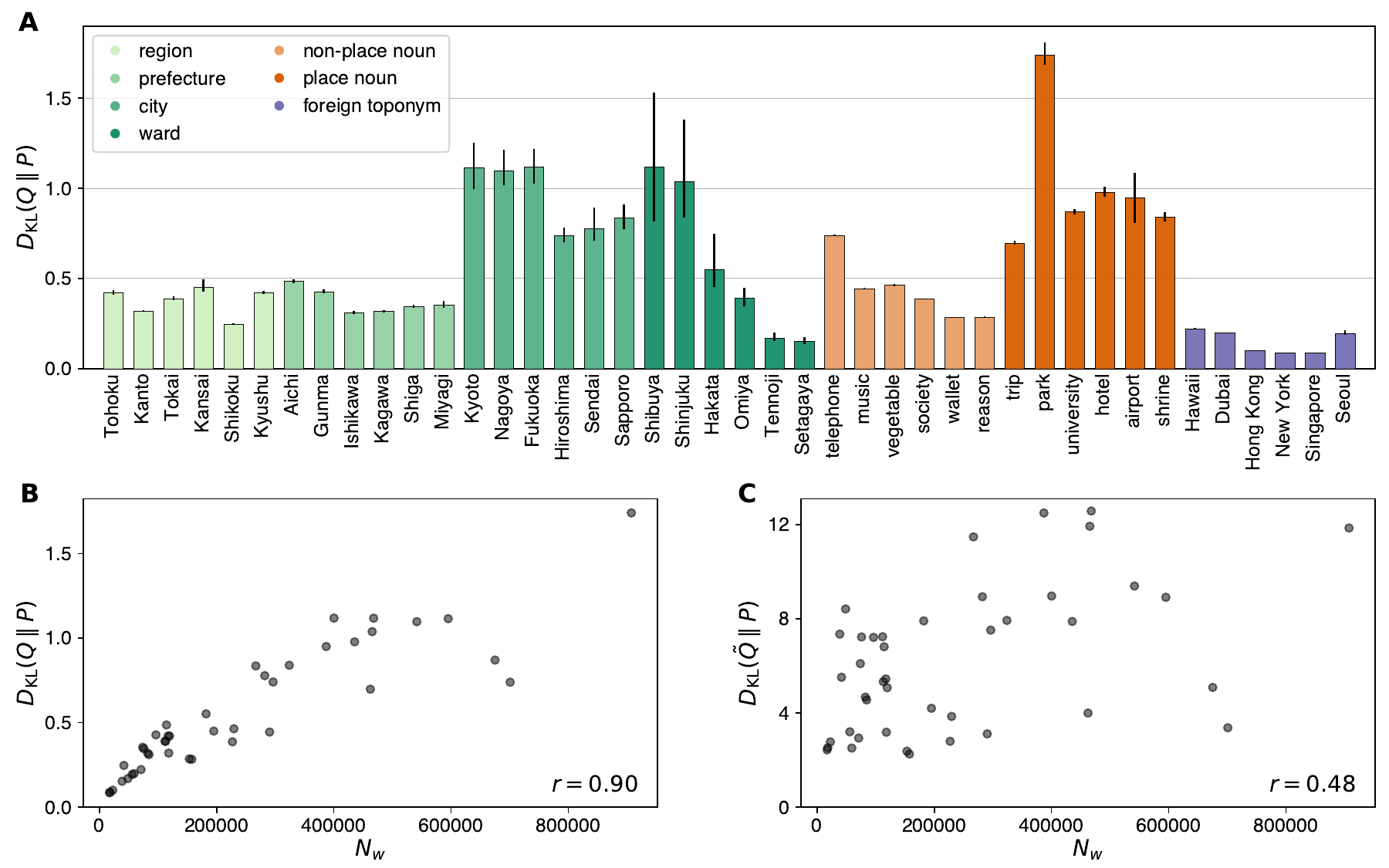}
    \caption{\textbf{Dissimilarity of empirical data from the location-independent model, evaluated by relative entropy.} \textbf{(A)} Relative entropy $D_\mathrm{KL} (Q_w \parallel P_w)$ for each toponym $w$ and noun studied in this work. For each word, 20000 grid cells are randomly sampled 50 times to compute the mean and 95\% confidence intervals, indicated by error bars. \textbf{(B, C)} Relative entropy $D_\mathrm{KL} (Q_w \parallel P_w)$ and the modified version $D_\mathrm{KL} (\tilde{Q}_w \parallel P_w)$, plotted against the total number of posts $N_w$ containing word $w$.}
    \label{fig:relative_entropy_vs_total_count}
\end{figure}

In Fig~\ref{fig:relative_entropy_vs_total_count}A, we present the estimated relative entropy for each toponym and each noun. As expected, toponyms denoting cities (such as ``Nagoya'') and some submunicipal districts (such as ``Shinjuku''), as well as place nouns (such as ``park''), are characterized by relatively larger values of relative entropy than those for common nouns without a place connotations (e.g., ``wallet''), implying that the empirical distribution is highly dissimilar from the location-independent binomial model. However, the relative entropies for toponyms that represent regions (e.g., ``Shikoku''), prefectures (e.g., ``Ishikawa''), and some other submunicipal districts (e.g., ``Setagaya'') are as small as those for common nouns.
This seems inconsistent with what we expect from Fig~\ref{fig:all_topo_many_lim} in Supporting Information, in which the difference between the data and the model is clearly visible for all domestic toponyms. This is presumably because the data contain more grid cells without any toponym occurrence, i.e., $n_w = 0$, than expected by the model. Such cells are prevalent even in cells with large $n_\mathrm{all}$, in which the model predicts a low probability for $n_w = 0$. The increase in relative entropy due to the presence of such zero-occurrence grid cells will be more pronounced for words with a large $\hat{p}_w$, or, equivalently, a large number of posts $N_w$. This hypothesis is supported by the strong positive correlation between $N_w$ and $D_\mathrm{KL} (Q_w \parallel P_w)$,  as shown in Fig~\ref{fig:relative_entropy_vs_total_count}B. Simply put, a large value of relative entropy may just be an artifact of the popularity of the word.

To discount the effect of inflated zero occurrences, we only sample grid cells with non-zero occurrences for computing the empirical distribution and relative entropy, i.e., $\Gamma \subseteq \{c \mid n_{w, c} > 0\}$. Let $\tilde{Q}$ denote the empirical distribution thus obtained. As shown in Fig~5 in the main text, this approach clearly discriminates the domestic toponyms and place nouns from nonspatial nouns and foreign toponyms.
The correlation between $N_w$ and $D_\mathrm{KL} (\tilde{Q}_w \parallel P_w)$ is much weaker (Fig~\ref{fig:relative_entropy_vs_total_count}C), further corroborating the hypothesis that the value of $D_\mathrm{KL} (Q_w \parallel P_w)$ is dominated by the contribution from zero-occurrence grid cells.

\section{Data preprocessing}
In this section, we provide the details of the data preprocessing procedures.
\subsection*{Selection of posts based on source applications}
As described in the main text, we limit the sample to posts sent from one of the official or general-use third-party mobile applications. Specifically, posts that have `source' metadata as one of the following applications are included:
`Twitter for iPhone',
`Twitter for Android',
`Instagram',
`Twitter for iPad',
`Twitter for Android Tablets',
`Path',
`Mobile Web',
`Mobile Web (M5)',
`Path 2.0',
`\japanese{ついっぷる} for iPhone',
`Photos on iOS',
`Twitter for Windows Phone',
`\japanese{ついっぷる} Pro for iPhone',
`Twitter for BlackBerry\textsuperscript{\textregistered}',
`Camera on iOS',
`\japanese{ついっぷる} for Android',
`\japanese{ついっぷる} Pro for Android',
`\japanese{ついっぷる} for Android org',
and `\japanese{ついっぷる} Pro for Android org'.
\mbox{\japanese{ついっぷる}} (Twipple) was a Twitter client provided by one of the major Japanese internet service providers, BIGLOBE, until it was discontinued in 2017.

\subsection*{Unnatural geotag distribution}
\begin{figure}[tbh]
    \centering
    \includegraphics[width=0.83\linewidth]{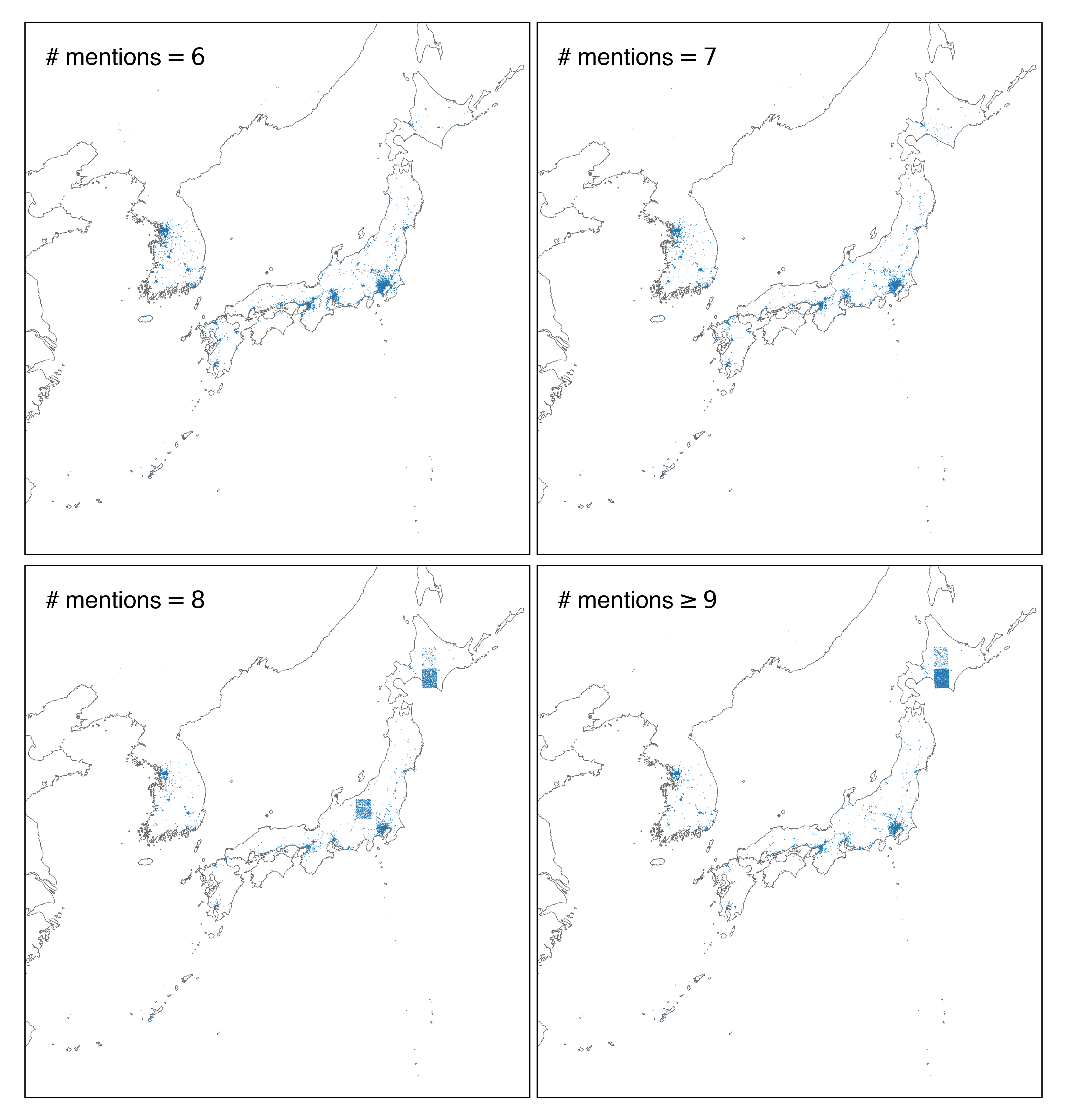}
    \caption{\textbf{Spatial distributions of posts containing a specific number of mentions.} Grid cells to which a non-zero number of posts are tagged are colored in blue.}
    \label{fig:spam_mentioncnt_map}
\end{figure}

During the data cleaning process, we noticed that the geotags attached to some posts had an unnatural geographic distribution, likely due to manipulation. Figure~\ref{fig:spam_mentioncnt_map} shows the geographic distribution of posts containing different numbers of mentions (references to other accounts). There are several rectangular blocks clearly visible in the distributions of posts containing mentions of eight other accounts and nine or more accounts. These blocks are unlikely to be the result of organic distributions for two reasons: (1) some of them extend over sea areas, and (2) such patterns are absent in posts with fewer than seven mentions.

\begin{figure}[tbh]
    \centering
    \includegraphics[width=0.42\linewidth]{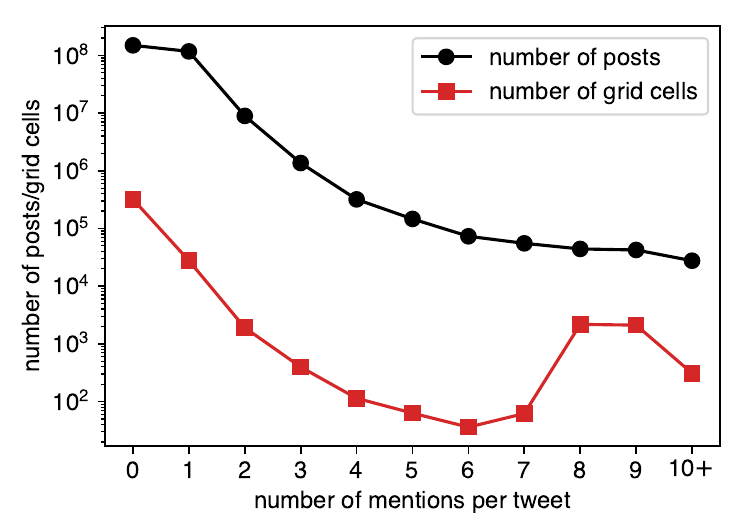}
    \caption{\textbf{Number of posts containing a specific number of mentions, and the number of grid cells to which they are tagged.}}
    \label{fig:spam_num_tweets_grids_by_mentioncnt}
\end{figure}

Figure~\ref{fig:spam_num_tweets_grids_by_mentioncnt} provides further evidence of manipulation. When posts are categorized by their mention count, the number of posts in each category decreases monotonically with the number of mentions. However, the number of unique grid cells where posts in each category are distributed shows an increase at a mention count of seven. Although the geographic distribution of posts with seven mentions does not exhibit visually clear symptoms of manipulation on the map (Fig~\ref{fig:spam_mentioncnt_map}), we opted to exclude all posts with seven or more mentions from the dataset as a conservative choice.

\subsection*{Toponyms that are substrings of other toponyms}
To exclude references to other toponyms from each toponym subsample, we identify the names of regions, prefectures, cities with a population larger than \num{50000}, and wards that contain each of the 24 toponyms we study in this work as a substring as follows:
\begin{itemize}
\item \japanese{関東} (Kanto) is a substring of \japanese{北関東} (Kitakanto) and \japanese{南関東} (Minamikanto).
\item \japanese{四国} (Shikoku) is a substring of \japanese{四国中央} (Shikokuchuo).
\item \japanese{九州} (Kyushu) is a substring of \japanese{北九州} (Kitakyushu).
\item \japanese{宮城} (Miyagi) is a substring of \japanese{宮城野} (Miyagino).
\item \japanese{京都} (Kyoto) is a substring of \japanese{東京都} (Tokyo-to, Tokyo Metropolis).
\item \japanese{名古屋} (Nagoya) is a substring of \japanese{北名古屋} (Kitanagoya).
\item \japanese{福岡} (Fukuoka) is a substring of \japanese{上福岡} (Kamifukuoka).
\item \japanese{広島} (Hiroshima) is a substring of \japanese{北広島} (Kitahiroshima) and \japanese{東広島} (Higashihiroshima).
\end{itemize}

\clearpage
\renewcommand{\thefigure}{S\arabic{figure}}
\setcounter{figure}{0}

\begin{figure}[ht]
    \centering
    \includegraphics[width=0.98\textwidth]{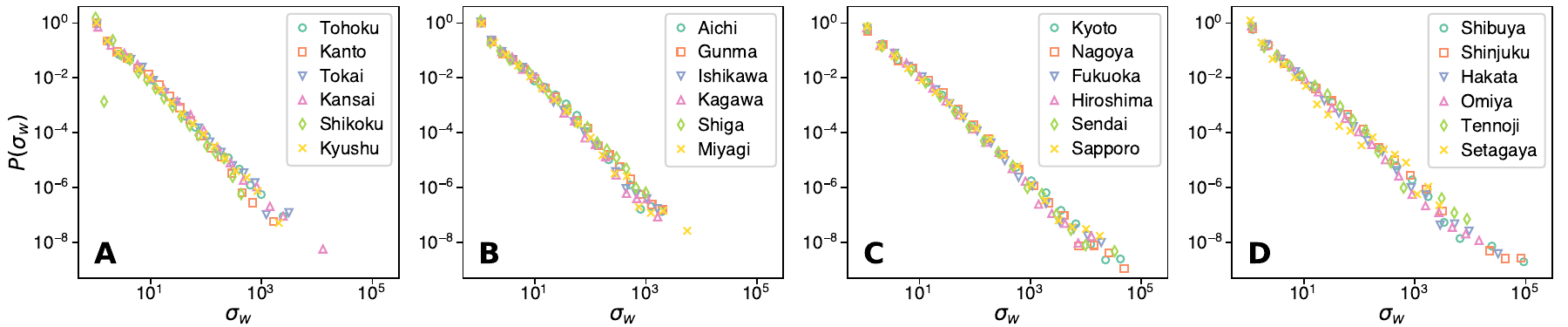}
    \caption{{\bf Probability density functions of occurrence density $\sigma_w$.}}
    \label{fig:density_pdfs}
\end{figure}

\begin{figure}[ht]
    \centering
    \includegraphics[width=0.98\textwidth]{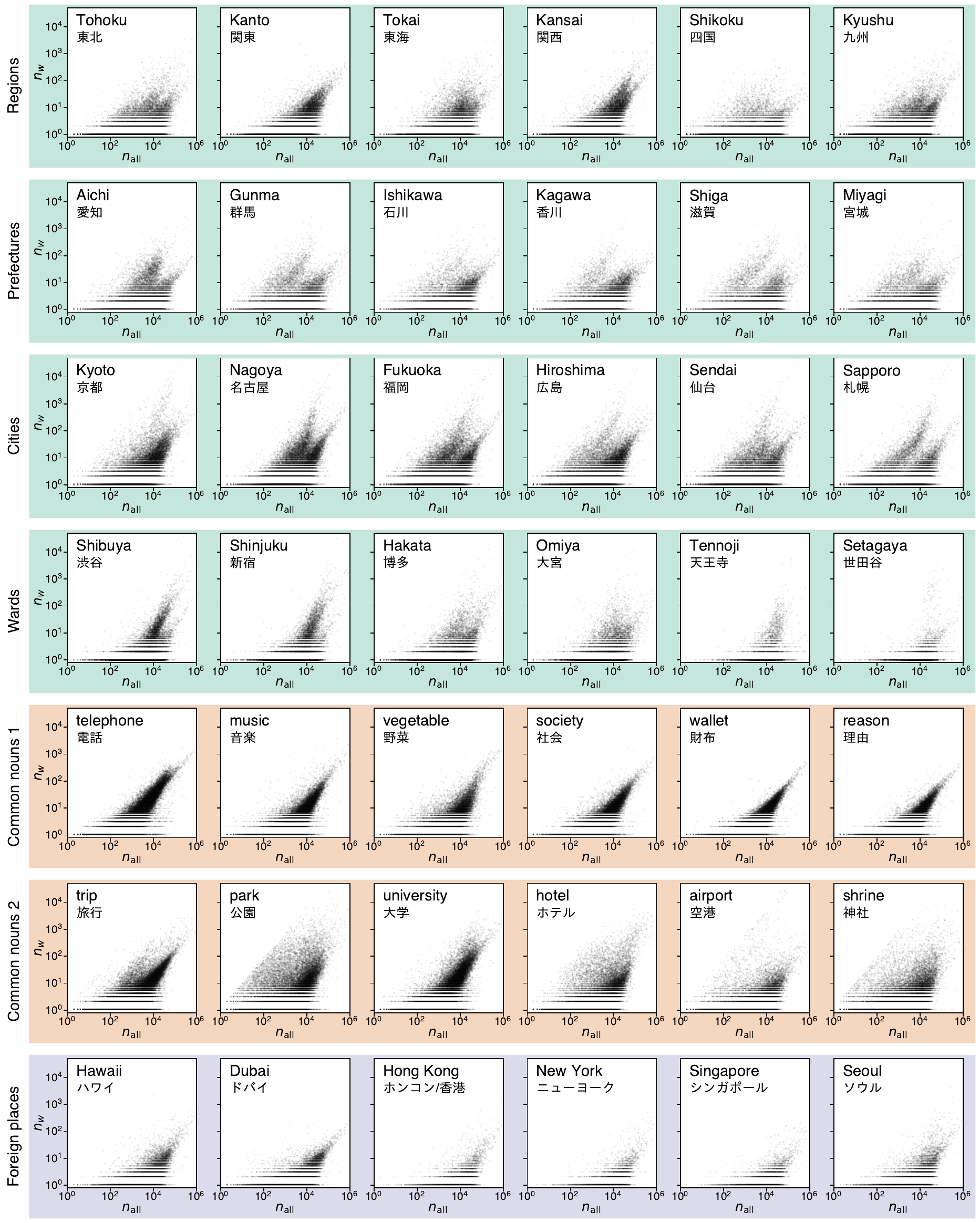}
    \caption{{\bf Scatter plot of $n_w$ versus $n_\mathrm{all}$ for all toponyms and nouns studied in this work.}}
    \label{fig:all_topo_many_scatter}
\end{figure}

\begin{figure}[ht]
    \centering
    \includegraphics[width=0.98\textwidth]{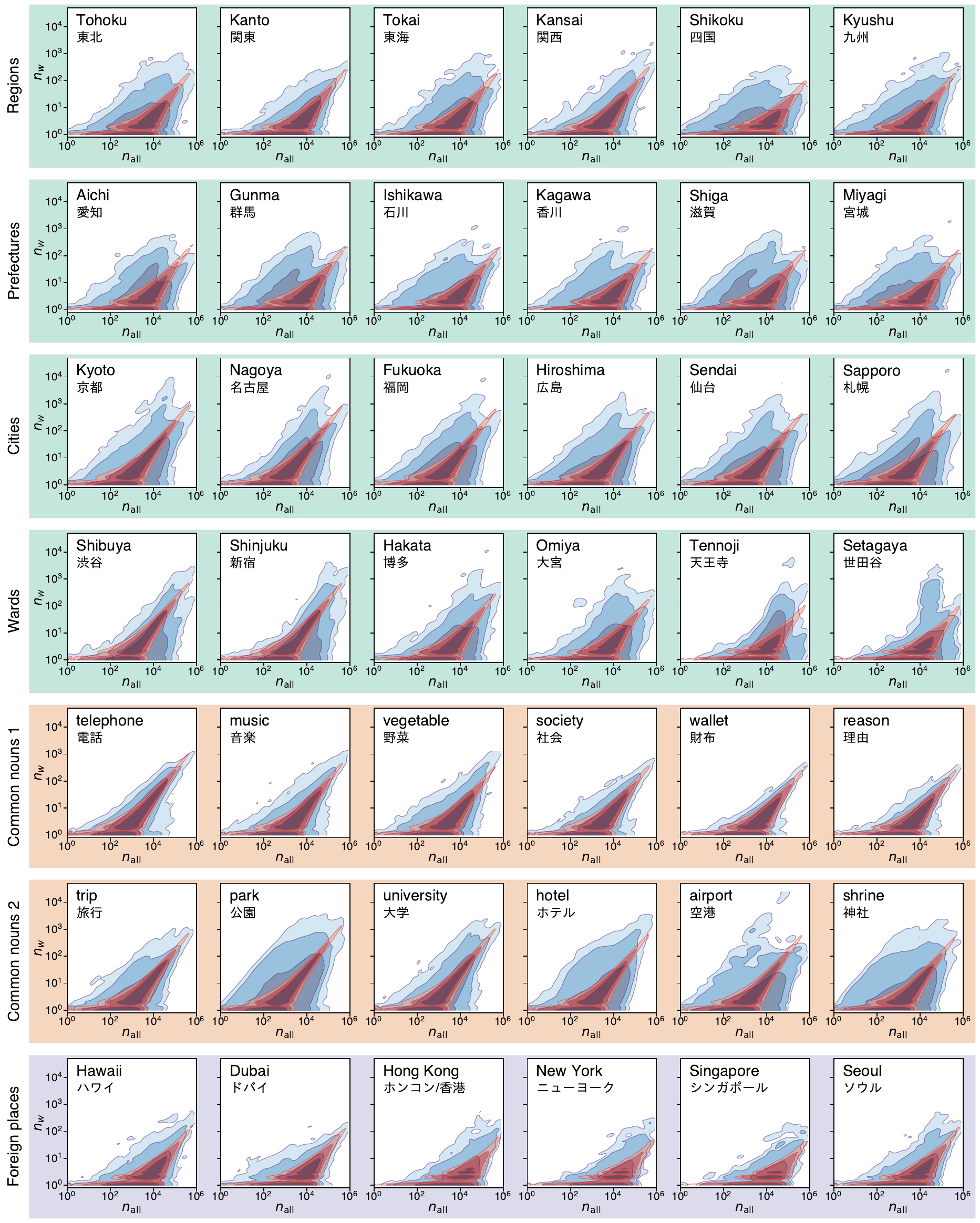}
    \caption{{\bf Comparison between the empirical data and the location-independent model.} Each set of contours represents the kernel density plot of $n_w$ versus $n_\mathrm{all}$ of empirical data (blue) and the location-independent model (red). }
    \label{fig:all_topo_many_lim}
\end{figure}

\begin{figure}[ht]
    \centering
    \includegraphics[width=0.98\textwidth]{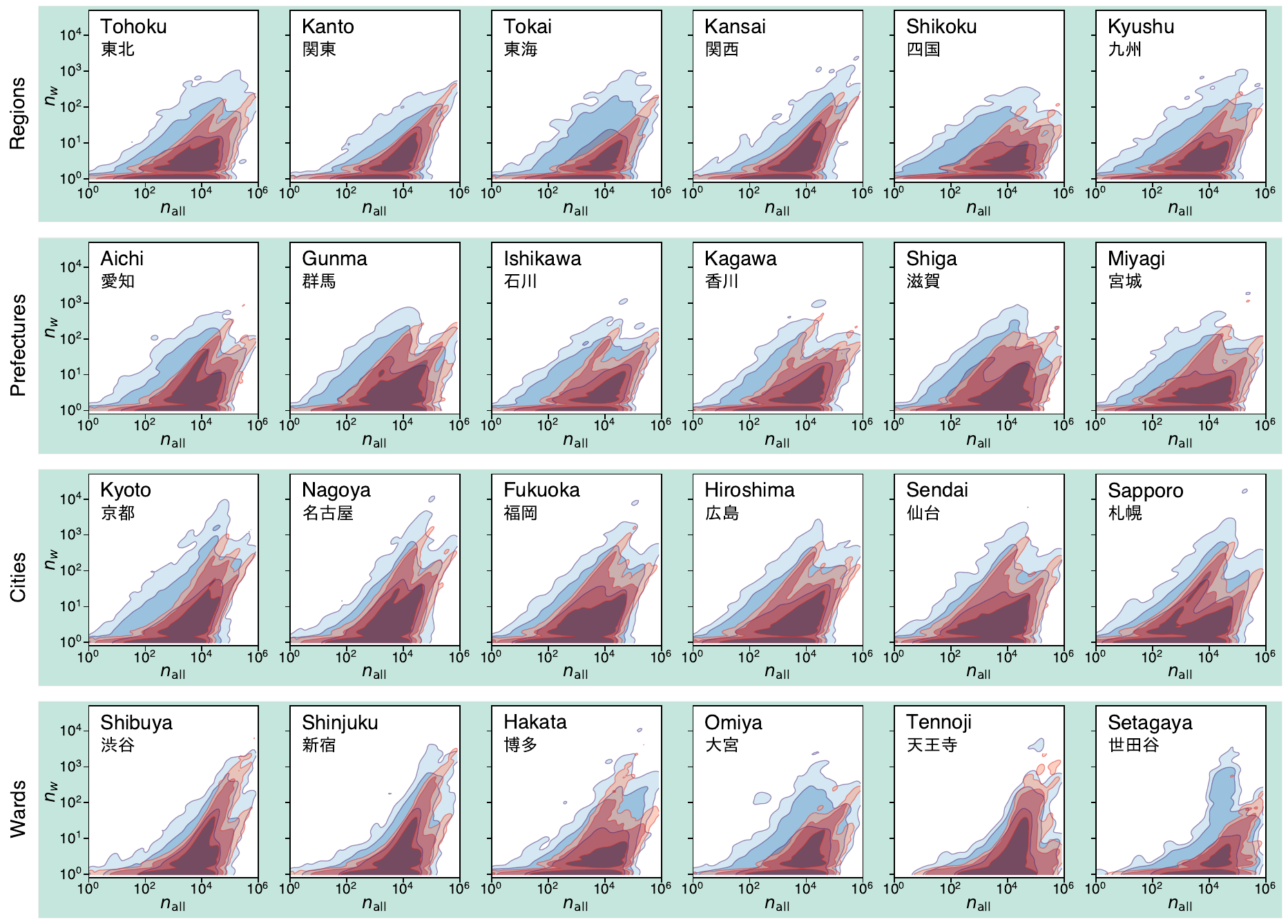}
    \caption{{\bf Comparison between the empirical data and the core-periphery model.} Each set of contours represents the kernel density plot of $n_w$ versus $n_\mathrm{all}$ of empirical data (blue) and the core-periphery model (red). }
    \label{fig:all_topo_many_cpm}
\end{figure}
\end{appendix}

\end{document}